\documentclass[twoside, epsfig]{article}
\input oejv.sty

\begin{document}

\OEJVhead{05 2022}
\OEJVtitle{Period changes of Mira variables in the M16-M17 region}
\OEJVauth{Nesci, Roberto$^1$; Soszy{\'n}ski, Igor$^2$; Tuvikene, Taavi$^3$} 
\OEJVinst{INAF/IAPS-Roma, via Fosso del Cavaliere 100, 00133, Roma, Italy {\tt \href{mailto:roberto.nesci@inaf.it}{roberto.nesci@inaf.it}}}
\OEJVinst{Astronomical Observatory, University of Warsaw, Al. Ujazdowskie 4, 00-478 Warszawa, Poland
 {\tt  \href{mailto:soszynsk@astrouw.edu.pl}{soszynsk@astrouw.edu.pl}}} 
 \OEJVinst{Tartu Observatory, University of Tartu, Observatooriumi 1, T\~oravere, Estonia, 
 {\tt \href{mailto: taavi.tuvikene@ut.ee}{taavi.tuvikene@ut.ee}}}

\OEJVabstract{We analyzed the light curves of 165 AGB variables, mostly Miras, in the sky area centered between M16 and M17 ($l=16$, $b=0$), using the OGLE GVS database in the $I_\mathrm{C}$ band. Comparison with the published light curves, derived about 50 years earlier by P.~Maffei using Kodak I-N photographic plates, allowed us to find no significant period changes in any star. 
Remarkably, a few stars of the sample appear to have substantially changed their average luminosity, the most striking case being KZ Ser.
We provide a better identification for three stars: IX Ser, NSV 10522, and NSV 10326, all of them being Miras. We classify the light curves of 6 stars, discovered but not classified by Maffei, (GL Ser, NSV 10271, NSV 10326, NSV 10522, NSV 10677, and NSV 10772) five of them being new Miras, and confirm the R CrB nature of V391 Sct. The magnitude scale used by Maffei is compared to the modern $I_\mathrm{C}$ one}.

keywords: Stars: AGB and post-AGB; Stars: late type


\begintext

\section{Introduction}\label{sec:intro} 

The red Long Period Variables (LPV) are stars on the Asymptotic Giant Branch, in evolution towards the planetary nebula phase. They are historically classified as Mira stars if they have a luminosity variation with a rather regular amplitude and a nearly constant period, or as semiregular (SR) if the amplitude and/or the period is not so regular.  Until recently there was no definite numerical threshold to distinguish between these two light curve shapes: a proposal in this sense was made by \citet{Sos13}. It is likely that in SR stars at least two periodic oscillations are present, with comparable amplitude, and that their interplay produces their less regular behavior.

An extensive search for the Mira period changes was made by \citet{Tem05}, based on the large database provided by the American Association of Variable Stars Observers (AAVSO). Out of 547 stars, 21 were found to have definite (greater than 3 $\sigma$) period changes in one century, the available  time extent of good quality light curves. Besides the well known case of T UMi, which is undergoing also a dramatic variation amplitude decrease \citep{Molnar19}, 7 stars showed secular trends larger than 30 days/century.
A similar work by \citet{Sabin06} based on the AAVSO and AFOEV (Association Francaise des Observateurs d'Etoiles Variables) datasets for 23 Miras with periods longer than 450 days, showed the existence of small period changes in most of these stars: generally the period variation is erratic, with differences in a few cases up to 10\% on a time interval of tens of years.

With the onset of robotic surveys of very large areas of the sky, very extensive databases  containing many kinds of variable stars have been produced. 
A work by \citet{Leb11}, based on 454 good quality light curves of Mira stars from the ASAS survey\footnote{\url{http://www.astrouw.edu.pl/asas/}},  discussed the shape of their light curves, finding about 30\% of the stars deviate significantly from a simple sinusoid.  Both asymmetric shapes and broad maxima exist.
The OGLE (Optical Gravitational Lensing Experiment) collaboration \citep{Uda15} has explored the behaviour of late type variable stars in the Magellanic Clouds \citep{Sos09, Sos11} and in the Galactic bulge \citep{Sos13}, providing good light curves in the $I_\mathrm{C}$ band, besides the classical $V$ band, and very tight period-magnitude relations. Relevant for this work is the finding that Mira variables have peak-to-peak $I_\mathrm{C}$-band amplitudes larger than 0.8~mag, while SR variables have smaller amplitudes.

In this paper we  search for possible period variations at 50 years distance in a set of 165 variables discovered by \citet{Maf75}, in the years 1967--75, in a 5$\times$5 degrees field centered at $l=16^\circ$, $b=0^\circ$. Finding charts and folded light curves of these stars were published only recently in \citet{Maf13}.
This field was covered recently by the OGLE Galaxy Variability Survey (GVS), which uses the 1.3-m Warsaw telescope located at Las Campanas Observatory, Chile.  
The GVS is a subproject of the OGLE survey, aimed at repeatable observations of the Galactic disk and bulge. Currently, the OGLE GVS project monitors brightness of more than one billion stars over an area of about 3000 square degrees. The telescope is equipped with a mosaic camera consisting of 32 CCD chips covering a field of 1.4 square degrees. A detailed description of the instrumentation, photometric reductions and calibrations of the OGLE data are provided in \citet{Uda15}.

The OGLE GVS observations were taken in the years 2014--2020, allowing to check possible variations in the light curves of our sample stars at a temporal separation of about 50 years. 
A period change can be easily recognized if the star is a regular Mira: uncertain Miras or SRs have by definition a variable period and/or amplitude. However we studied all the stars of the sample to check for possible misclassifications or actual transitions from Mira to SR and vice versa.

Furthermore,  attempting to merge the old Maffei's and the new OGLE datasets, we tried to transform the magnitude scale used by \citet{Maf13} into the modern $I_\mathrm{C}$ OGLE magnitudes.

We remark that, for the large majority of our stars, no further studies were published after the discovery papers.

\section{Historic Maffei's Observations}\label{sec:obs} 

The variable stars of this sample were discovered by \citet{Maf75} using the Newtonian 122 cm and 65/92 cm Schmidt telescopes of the Asiago Observatory between 1961 and 1975 using photographic plates of I-N emulsion and RG5 filter, fairly reproducing the $I_\mathrm{C}$ band (see Section \ref{sec:comp-mags}). A blue plate (103a-O emulsion) of the field was always taken by Maffei in the same night for comparison, but only a few of these variables were detected also in the blue band, as expected for very red stars.

The periods of these variables were later refined by Maffei, using plates with a nearly identical filter/emulsion combination, taken with the 40/50 cm Schmidt telescope of the Catania Observatory in the year 1980,  with a few plates taken in 1983--84, and 1990--91. The finding charts, phase-reduced light curves, periods, epoch of maximum, observed maximum and minimum brightness, were finally published by Maffei and Tosti in 1999 and later put into a digitized catalog \citep{Maf13}, available at CDS (Vizier II/320). The number of photometric points was not the same for all stars, ranging from 49 to 143: in this respect the stars belong to two groups: one of 87 stars with a median of 57 observations, and one of 67 stars with a median of 122 observations.
Besides the published papers, we found in Maffei's private library also two unpublished Thesis works (supervisor Maffei) regarding these stars: one by Mario Montalto (La Sapienza University, 1971) and one by Sergio Schioppa (Perugia University, 1983), which  permitted correction of some misprints in the periods and light curve classifications.  Finally, we found also the tables of the original magnitude measures, which allowed us to perform new period determinations with the same software used for the analysis of the OGLE data.

The variability type was assigned by Maffei on the basis of the light curve shape and amplitude. 
Overall the stars in the sample were classified as follows: 8 eclipsing stars (E or EA), 2 irregular (I), 111 Mira (M), 14 uncertain Mira (M:), 24 semiregular (SR). For 7 stars Maffei did not give a classification. For 15 stars (including the unclassified ones, the irregulars and most eclipsing stars) Maffei did not give a period. The periods were derived with the \citet{Press89} software based on the \citet{SC96} method: accuracy for well sampled stars is quoted as a few days in Montalto's thesis. Some periods however are indicated as uncertain (marked with colon (:) in  Vizier II/320 Table 3).

The coordinates of many stars of our sample were checked by \citet{Kato01} by comparison with the MSX5C catalog. Later, one of us \citep{Nes18} checked the coordinates of all the stars comparing the original Maffei's finding charts with the UK Schmidt IV-N plates and the 2MASS catalog, allowing the correction of a few misprints in the original papers and a better identification of some variables;  for three stars we further improved the identification during this work (see Section~\ref{sec:remarks} below).

\section{Modern OGLE observations and Data analysis.}\label{sec:data}
\subsection{Periods of the OGLE light curves.}\label{sec:det-per}

Light curves for 165 Maffei's variables were found in the OGLE GVS database, with a typical sampling for each star of about 100 points over a 5-year time span, comparable to that of Maffei's light curves. The photometric errors were  usually smaller than 0.01 mag.
A few known eclipsing variables were excluded, as well as a few very poorly sampled stars, giving a final sample of 150 stars. 

The periods of these variables and 1$\sigma$ errors were measured with the Tatry code, which implements the multiharmonic analysis of variance algorithm \citep{SC96}. The median error of our periods estimates is 0.76 days, with standard deviation of the distribution 0.66 days.

We computed the epoch of maximum and variation amplitude with a simple sinusoidal fit using the Period04 code \citep{Lenz05}, assuming the period given by the Tatry code. The epoch of maximum was used to test the predictions of Maffei's ephemeris, and the variation amplitude to check for possible major changes in the light curve behaviour, keeping in mind the case of T UMi.

While our paper was nearly finished, the OGLE team published a large catalog of Mira variables found in the galactic bulge and disk, including most of our stars \citep{Iwa22}. The periods and amplitudes given in this catalog are consistent with ours within the observational uncertainty: the few discrepant cases will be discussed in the next Section. 

In Table 1 we present for each star the SIMBAD name, the 2MASS counterpart, the identifier from the OGLE Collection of Variable Stars (OCVS), Maffei's number, our period, the formal error, the epoch of maximum, the observed maximum and minumum magnitudes rounded at 0.1 mag. For the stars not present in the OGLE catalog we give the internal identifier of the light curve in OGLE database. A Period=0 or err=0 means no actual measure for lack of enough data. No error is also given for the very short-period variables found.

The full Table 1 in ASCII format is reported after the references.

\begin{table*}
\caption{Light curve parameters from OGLE data}\vspace{3mm}
\centering
\begin{tabular}{l l l l r r r r r}
\hline
SIMBAD & 2MASS & OGLE & Maffei  & Per & errP & Epoch& Max& Min \\
ID     &   ID  & ID   &   ID    &days & days & MJD  & mag & mag   \\
\hline
\hline
FZ Ser&18080193-1444151&OGLE-BLG-LPV-258610&M173&243.1&0.4&56939&12.5&15.7\\
GG Ser&18081103-1434279&OGLE-BLG-LPV-258664&M101&362.1&0.8&56879&10.8&12.2\\
GH Ser&18082202-1524048&OGLE-BLG-LPV-258711&M030&276.3&0.5&56917&12.1&14.2\\
GI Ser&18082639-1535119&OGLE-BLG-LPV-258733&M090&252.1&0.5&56963&11.2&14.0\\
GK Ser&18082549-1418077&OGLE-BLG-LPV-258729&M103&373.1&0.4&57123&13.6&16.1\\
..... & .............. & ............  & ...& ....  & ....& ... & ...& ..\\
\hline
\end{tabular}\label{table1}
\end{table*}

We remark that for stars with period around one year (350--380~d, about 20 stars) there may be a period ambiguity between one year and half year due to the seasonal gap, which is rather large despite the telescope location in the Southern Hemisphere: this gap is even larger in the historic Maffei's data.

\begin{figure*}
  \includegraphics[width=15cm]{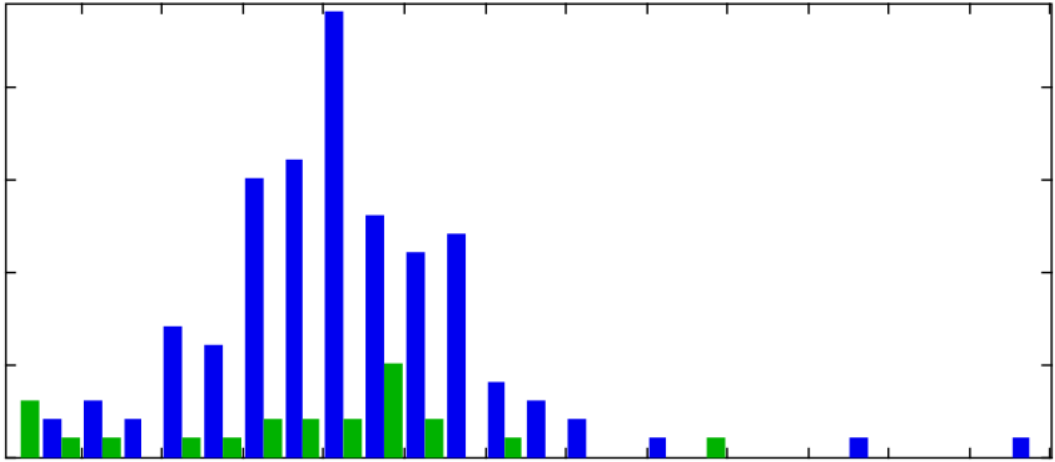}
  \caption{Distribution of variability amplitude based on the OGLE light curves for stars classified by Maffei as Miras and SR variables.}
  \label{fig:histogramsA}
\end{figure*}

The histogram of the derived amplitudes for the Mira and SR stars is given in  Fig.~\ref{fig:histogramsA} and shows no difference between these two classes, so that it cannot be used to discriminate between them. The puzzling result of this plot is that nearly all stars have amplitude larger than 0.8~mag in $I_\mathrm{C}$, including those classified by Maffei as SR, suggesting that all of them should actually be considered as Miras according to the simple criterion of variation amplitude $\le$ 0.4~mag \citep{Sos13}.

\begin{figure*}
  \includegraphics[width=15cm]{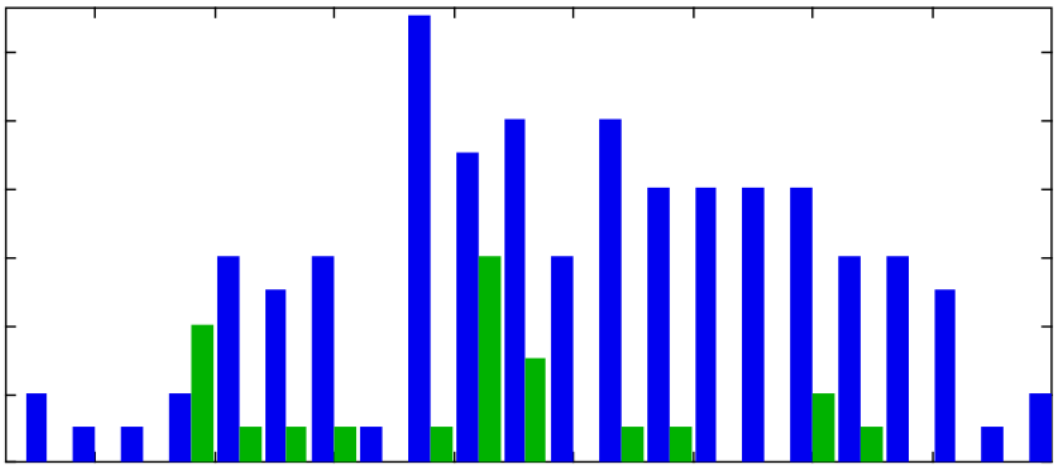}
  \caption{Distribution of periods based on the OGLE light curves for stars classified by Maffei as Miras and SR variables.}
  \label{fig:histogramsP}
\end{figure*}

The periods histogram (Fig.~\ref{fig:histogramsP}) shows a little prevalence of longer periods for the stars classified by Maffei as Mira, but the overlap of the two distributions is quite large, again suggesting a revision of the classification.

\subsection{Comparison of Maffei's and OGLE periods.}

\begin{figure}
\centering
\includegraphics[width=15cm]{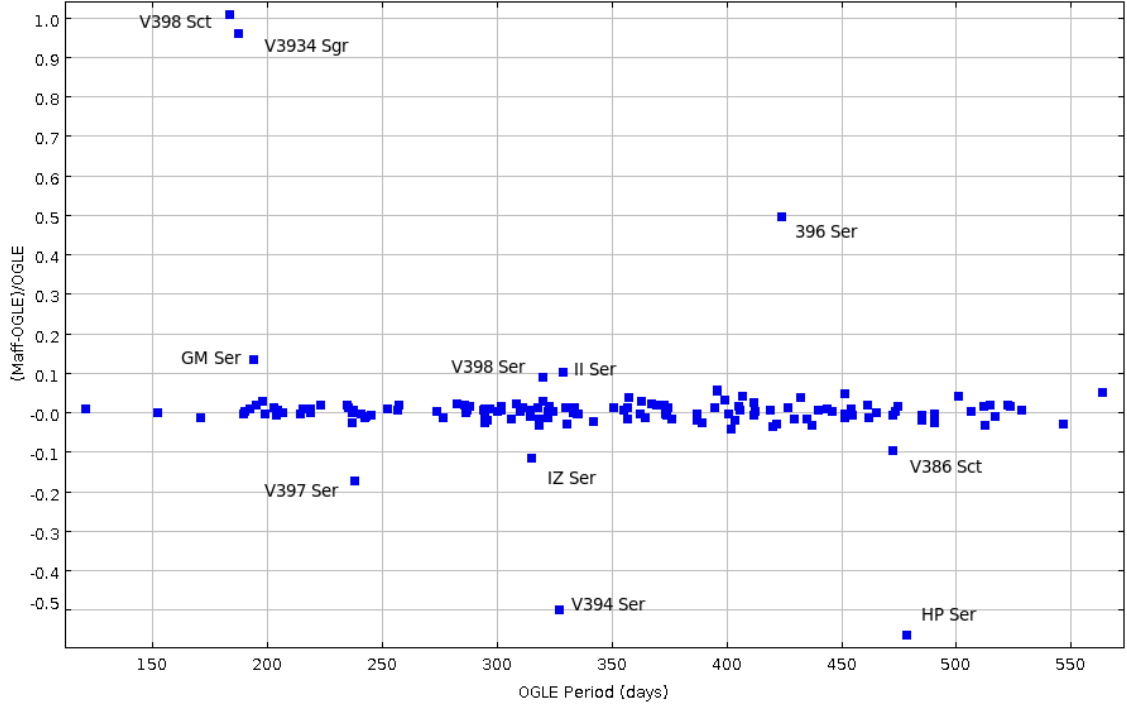}
\caption{Ratio of the differences between Maffei's and OGLE period to the OGLE period ($\Delta P/P$), vs the OGLE period. Stars that exhibit a large period discrepancy are labelled and discussed in the text.}
\label{fig:periods}
\end{figure}

Having derived the present periods of our stars from the OGLE GVS database, we plot in Fig.~\ref{fig:periods} for each star the ratio of the difference between Maffei's and OGLE period to the OGLE period ($\Delta P/P$): most of the stars show very consistent results, indicating that no significant period changes happened between the two observation campaigns.

Having found in Maffei's archive his original magnitude measures, we used our Tatry code to derive new period determinations from the old Maffei's observations, to check the effect of using a different code on the period determination: these new period measures were largely consistent with the original ones. Indeed, for a subsample of 129 stars the average difference between the two measures was just 0.21 days (and the median difference even smaller, 0.09 days); the rms scatter was 2.88 days and the maximum period difference found was 13 days, i.e. 5$\sigma$. 
For 6 stars the difference was about a factor 2 or 3, suggestive of period aliases taken as actual periods.
For 8 stars the difference was larger than 20 days, anyway below 14\% of the period. In 4 cases the Tatry code did not converge to give a Period.

Another way to look for period changes, besides the simple numerical comparison, is to check if the OGLE ephemeris (period and epoch of maximum) predicts correctly the date of Maffei's epoch (or vice versa). The number of cycles for our variables between the years 1967 and 2014 ranges from about 30 (for a 500 d period) to 70 (for a 200 d period), so that an error of one day (the typical OGLE error) in the period produces a difference of about one to two months in the computed epoch, and a phase difference ranging from 0.06 to 0.28. The accuracy of Maffei's epochs is not declared, and we assume it should be about one week from the shape of a typical Mira light curve. 

We made this test computing the expected phase of each light curve at Maffei's epoch, finding a nearly flat distribution between -0.5 and 0.5 instead of a clustering around phase zero as expected if the periods were nearly constant. The same happened computing the expected phase of the light curve at the OGLE epoch using Maffei's ephemeris.

However, given the large number of cycles between the two epochs, a small difference in the adopted period could easily adjust the phase to the nominal value.
Indeed we computed for each star an average period, using as time base the interval between Maffei's and OGLE epochs of maxima, and a number of cycles nearest integer to that derived from the OGLE period. 
Given the very large time base (of the order of 16400 days) and the 365 days length of a typical period, the formal accuracy of these average periods would be about 0.01 days, even if the epoch uncertainty is assumed 2 weeks.
If the period were stable, the period computed in this way should be very similar (within a few $\sigma$) to the OGLE one, measured on a 4 year time base.

We made this test, finding periods differences $\le$ 3$\sigma$ for 111 stars, $\le$ 4$\sigma$ for further 10 stars, and $\ge$5$\sigma$ for 14 stars.

As told in the Introduction, fluctuations of a few percent are rather common in Mira variables over tens of years, so we will discuss in detail only the apparently very discrepant stars.

For the large majority (104/111) of the Mira stars the period percent difference ($\Delta P/P$) between OGLE and Maffei was less than 0.04, with rms deviation 0.02, indicating that no significant changes happened in 50 years: only two Miras showed a period discrepant by more than 5\%.

Out of the 14 Maffei's probable Miras (M:) only one showed a significant difference. 

Out of the 24 SR, two do not have a period by Maffei; three showed a marked period variation ($\Delta P/P \ge 0.5$), whereas three others have $\Delta P/P$ about 0.1. It is remarkable that so many (6/24) stars classified by Maffei as SR have a significantly different period, observed 50 years later, compared with the (2/111) Miras. This may suggest that the criterion to discriminate Miras and SRs just from the light curve amplitude may not be completely sufficient.

Finally, Maffei labelled 8 stars as eclipsing variables, but only 4 have a period reported, always very long, ranging from 196 to 592 days. From their OGLE light curves these four stars are classified as follows: GM Ser is an R CrB star; V374 Sct is a Mira with a period similar to Maffei's one; V382 Sct has only 50 points in OGLE but looks rather irregular;  V397 Ser looks a SR.
Of the remaining four stars, two are actually short-period eclipsing stars (NSV 10497, $P=4.11316\,\mathrm{d}$; NSV 10681, $P=1.84228\,\mathrm{d}$); one is a Mira (NSV 10249, $P=191.5\,\mathrm{d}$); one looks a SR (NSV 10832, $P=320\,\mathrm{d}$).

\subsection{Discrepant stars}\label{sec:discrepant}

\begin{figure*}
  \includegraphics[width=\linewidth]{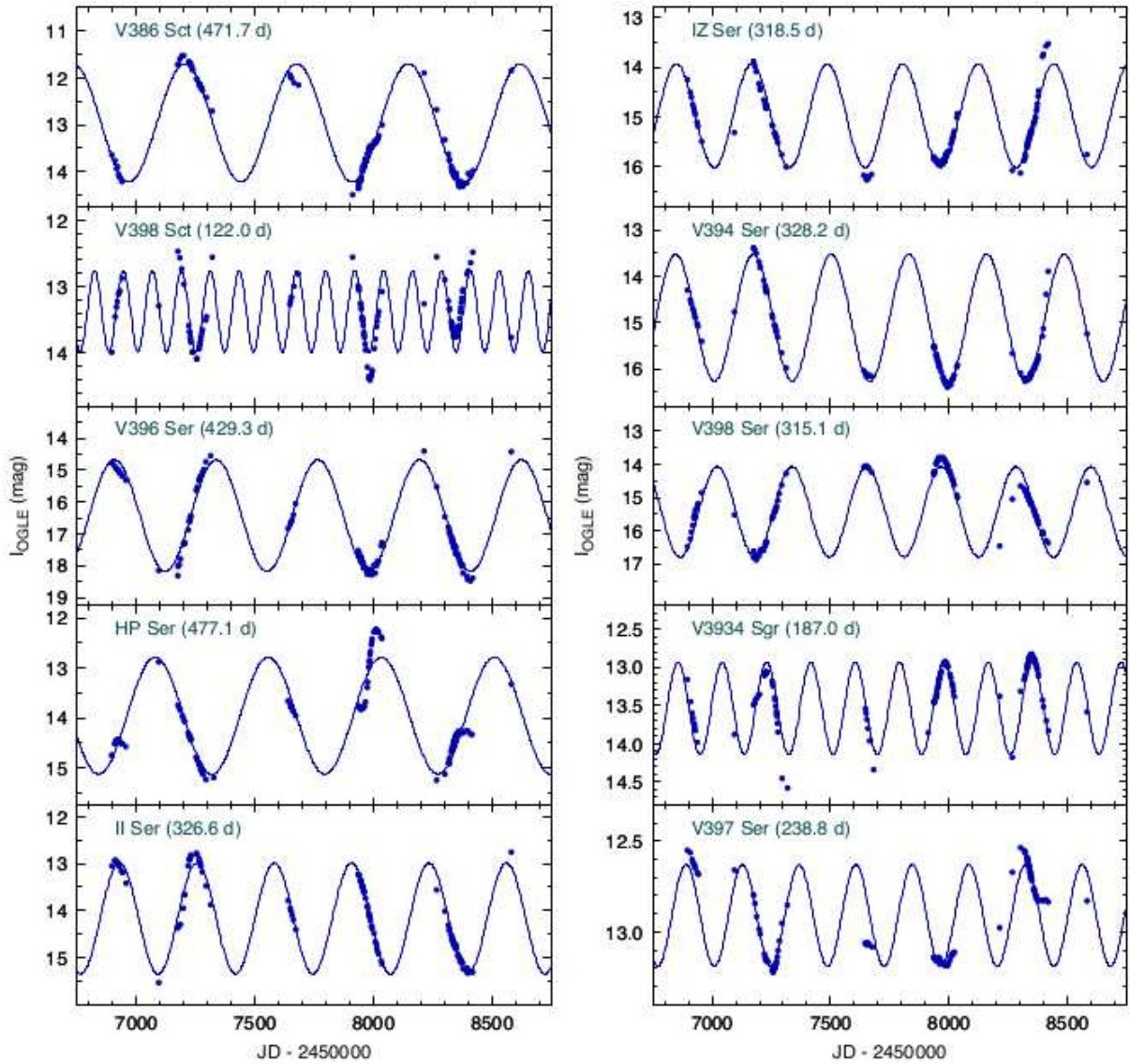}
  \caption{OGLE photometry with fitted light curves  for the period-discrepant stars.}
  \label{fig:lightcurves}
\end{figure*}

The most period-discrepant stars are listed in Table~\ref{tab1a}, grouped according to the original Maffei's classification, and their OGLE light curves are shown in Fig.~\ref{fig:lightcurves}.  We report in column 1 the SIMBAD name; in column 2 Maffei's number; in column 3 the OGLE identifier; in column 4 Maffei's classification; in column 5 Maffei's period; in column 6 the OGLE period; in column 7 the formal OGLE period error; in column 8 the ratio of the periods difference to the OGLE period ($\Delta P/P$).

We examined more in detail these stars to see if they really underwent a period change, and are discussed below. The interactive code Peranso \citep{Pau16} was used to interactively build phased light curves with different periods and evaluate the significance of the adopted periods.

\begin{table*}
\caption{Stars with discrepant periods}\label{tab1a}\vspace{3mm}  
\centering
\begin{tabular}{l l l l r r r r}
\hline
Name&Maffei&OGLE& Cl.& $P_\mathrm{Maffei}$ & $P_\mathrm{OGLE}$&errP&$\Delta P/P$ \\
 & & &  & d & d & d &  \\
\hline
\hline
V386 Sct&M026&OGLE-BLG-LPV-263898&M&426&472.4&1.0&-0.097\\
V398 Sct&M145&OGLE-BLG-LPV-264150&M&368&183.3&0.4&2.016\\
\hline
V396 Ser &M132&OGLE-BLG-LPV-261077&M:&633:&423.9&1.41&0.474\\
\hline
HP Ser &M093&OGLE-BLG-LPV-259581&SR&208&478.1&1.1&-0.564\\
II Ser  &M041&OGLE-BLG-LPV-259900&SRa&362&328.7&0.8&0.108\\
IZ Ser  &M178&OGLE-BLG-LPV-260964&SR&278&314.8&1.1&-0.127\\
V394 Ser &M046&OGLE-BLG-LPV-260384&SR&163&326.9&0.4&-0.503\\
V398 Ser &M186&OGLE-BLG-LPV-264150&SR&348&319.9&1.5&0.104\\
V3934 Sgr &M109& OGLE-BLG-LPV-262824&SRa&366&187.0&0.4&0.957\\
\hline
V397 Ser &M175&BLG780.16.28126 &EA&196&237.8&0.7&-0.179\\
GM Ser   &M092&BLG766.05.11    & E&219&193.8&0.8&0.132\\
\hline
\end{tabular}\label{table2}
\end{table*}

a) Mira stars.

V386 Sct: its period is indicated by \citet{Maf13} as 426 d in their Table 3, but as 446 d in their Figure of the phased light curve. We re-measured Maffei's period with the Tatry code using Maffei's magnitudes finding 447 d, in fair agreement with the OGLE one. The OGLE light curve is nicely sinusoidal with a small rms deviation and a period of 472 d, so it does not seem a case of significant period change.

V398 Sct: the OGLE light curve shows minima of different depth and a period of half year, while Maffei gives one year. The phase-reduced light curve published by Maffei shows a large spread around the maximum phase. Maffei's data are not incompatible with the OGLE period (183 d): likely Maffei's estimate was biased by the seasonal gap and a real period change is not likely.

V396 Ser (NSV 10472): The light curve in \citet{Maf13} has only 10 points and the quoted period is tentative. The light curve in the OGLE database is well populated and rather regular, with a large amplitude (4.0 mag peak to peak) and a period not compatible with Maffei's suggestion. It can be classified safely as Mira, but it is not possible to check if a period change really happened.

b) SR stars.

HP Ser: the OGLE light curve shape shows a strong bump in the ascending branch, not uncommon in Miras, with variable level. Maffei's period recomputed with Tatry is 492 d, similar to the 478 d OGLE one, with a difference of just 3\%; the variable amplitude of the light curve, and the low accuracy of Maffei's magnitudes makes the 14 days difference not very significant.

II Ser: the OGLE light curve is quite regular, typical of a Mira. Maffei's period is not compatible with the OGLE light curve, but his phased light curve is bad. Our recomputation of Maffei's period with Tatry code gives 3 possible periods of nearly equal weights, one of them similar to the OGLE one: there is no evidence therefore of a period variation.

IZ Ser: the OGLE light curve is well behaved and Mira-like but with variable amplitude. Our recomputation of Maffei's period with Tatry gives the same period of the OGLE data. We remark that using the Lombe-Scargle method on Maffei's original data gives the same result given by Maffei, while for the OGLE data gives nearly the same period of the Tatry code. This difference is likely due to the much larger photometric errors of the photographic magnitudes. Also for this star there is no evidence of period variation.

V394 Ser (alias NSV 10391); the OGLE light curve is well behaved and with a period double of Maffei's one. The phased light curve by Maffei shows several points at odds. Our reanalysis of his data shows several periods of nearly equal weight (164, 226, 328 d), the shortest one being Maffei's one, and the longest one being the OGLE one. Maffei's data are strongly dominated by a dense sampling, 170 days long, made in 1980. Mayby the star was correctly classified as SR by Maffei, but its behavior was quite regular during the OGLE monitoring.

V398 Ser (alias NSV 10479): Maffei's light curve has a small amplitude due to many upper limits; the OGLE period is definitely different from Maffei's one. It shows a large amplitude variation and small phase shifts, so it is likely a Carbon star, not a regular Mira.

V3934 Sgr: the OGLE light curve shows large amplitude variations between consecutive cycles. Maffei's data can be well fitted by the OGLE period or by its double with very similar significance, so there is no convincing evidence of a transition from the fundamental frequency to the first overtone.

c) eclipse star:

V397 Ser (alias NSV 10475): Maffei does not show a phased light curve of this star, classified as eclipse, and its period is flagged uncertain. The OGLE light curve shows a small amplitude (0.6 mag peak to peak) and a quite scattered phased light curve using a 238 d period. The star looks an Irregular or SR variable.

GM Ser: classified as eclipse by Maffei with spectral type M7 (Montalto's thesis); classified as CrB in SIMBAD. The OGLE light curve is well fitted by a small-amplitude sinusoid, but shows a sharp deepening and a small flare, indicative of an R CrB: therefore its light curve is not shown in Fig.~\ref{fig:lightcurves}. \citet{Tis13} have contested its R CrB classification,  but confirmed the spectral type M7.

\subsection{Further remarks}\label{sec:remarks}

{\bf Wrong identifications.}

During the present work we found three stars with wrong identification.

NSV 10326 (Maffei M005): no period and classification were given by Maffei for this star. The present identification in SIMBAD is with a blue star, much brighter than the value reported by Maffei; we found a source, at 20 arcsec distance, 2MASS J18110050-1423082, with quite red colors, whose OGLE light curve is typical of a Mira star with a period of 315 days. We therefore identify this star with  Maffei's variable M005. 

NSV 10522 (Maffei M035): no period and classification were given by Maffei for this star. The SIMBAD counterpart of M035 was found substantially constant in the OGLE data. We made therefore a new careful comparison of Maffei's finding chart with the DSS I-band plate available from the STScI and the 2MASS image from SIMBAD, finding a new candidate at 9 arcsec distance, 2MASS J18182915-1725379. The OGLE light curve showed that this star is a large amplitude Mira with a period of 457 days.

IX Ser (Maffei M179): our identification is with 2MASS J18153810-1505332, 9 arcsec off the quoted position in SIMBAD. Its OGLE light curve is typical of a Mira variable, with amplitude and period very similar to those quoted by Maffei (amplitude 1.66 mag, P=303 d).

{\bf Other variables.}

Two stars were classified as irregulars by Maffei and therefore had no period measure. One of them, NSV 10899 (M024, 2MASS J18295552-1518384), has a time scale of 34 days in the OGLE data and we classify it as an RV Tau star. In the OCVS catalog its period is reported as 68 days, double than ours. The second star, V391 Sct, has an R CrB light curve and was already classified as such by \citet{Tis13}. 

{\bf Unclassified stars.}

For six unclassified stars of Maffei's sample we derived the light curve and the period from the OGLE database: we have checked the VSX database to see if they have been studied in detail after Maffei's publications, finding nothing. Five of them are Mira variables, both from the amplitude and from the period length, and one is a SR; two stars are new identifications, as discussed above.
Our measures of periods, amplitudes, mean magnitudes, and classifications for these six stars are listed in Table~\ref{tab3}.

\begin{table*}
\caption{OGLE data for Maffei's unclassified stars}\vspace{3mm}  
\centering
\begin{tabular}{l l l r r r l}
\hline
Name & Maffei & OGLE & Period & Amp.& mean I$_C$ & Class.\\
     &        &      &   d    & mag & mag &   \\
\hline
\hline
GL Ser   & M031 & OGLE-BLG-LPV-258777 & 171 & 1.23 & 13.70 & Mira \\
NSV 10271& M174 & OGLE-BLG-LPV-258970 & 300 & 1.73 & 14.89 & Mira \\
NSV 10326& M005 & BLG766.18.154       & 315 & 0.93 & 15.31 & Mira \\
NSV 10522& M035 & OGLE-BLG-LPV-261605 & 456 & 1.97 & 16.19 & Mira \\
NSV 10677& M028 & OGLE-BLG-LPV-262404 & 394 & 1.13 & 15.32 & Mira \\
NSV 10772& M183 & BLG569.19.104       & 262 & 0.12 & 13.74 & SR  \\
\hline
\end{tabular}\label{tab3}
\end{table*} 

{\bf Corrections of mistakes.}

During this work we found two inconsistencies in \citet{Maf13}, which are listed below.

M015, V405 Sct: the period is 307~d in Table~3 but 357~d in the light-curve Figure; in the discovery paper \citep{Maf75} the period was 362~d. The OGLE value 359.2~d is in agreement with the original value.

M203, V374 Sct: originally classified as SR by \citet{Maf75} with 400~d period, and in the thesis work by Schioppa (1983) with 408~d period, from a dataset of 38 points. Later it was reclassified as E in \citet{Maf13}. The OGLE data (110 points) show a Mira variable with 420~d period. Our reanalysis of Maffei's data with Tatry confirms a 409$\pm$2~d period: the difference is only 2.6\%, not much relevant.
 
Overall, after our careful analysis,  no star among those classified as regular Miras appears to have changed its period by a significant amount. 

\section{Comparison of the OGLE magnitudes with Maffei's ones.}\label{sec:comp-mags}

\begin{figure}
\centering
\includegraphics{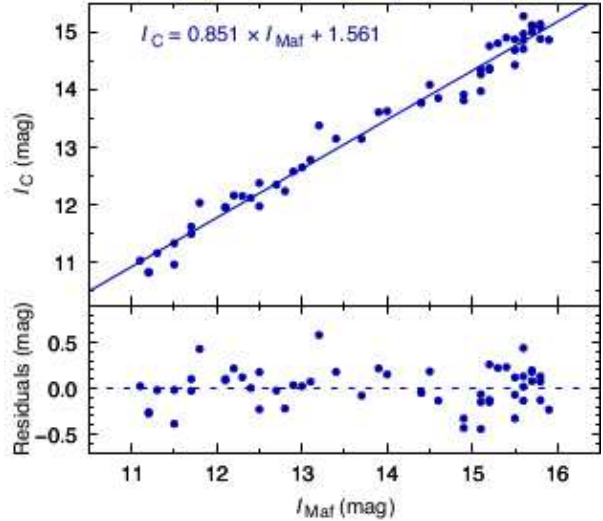} 
\caption{Comparison of the original Maffei's $I$ magnitudes of the comparison sequence with the OGLE $I_\mathrm{C}$ magnitudes.}
\label{fig:mag_calib}
\end{figure}

After checking the periods, we tried to verify if any star had significantly changed its average magnitude or variability amplitude.
 
As written in Section \ref{sec:obs}, the filter and emulsion used at the Asiago Observatory to obtain infrared plates were repectively RG5 and Kodak I-N. The RG5 filter is a longpass with cutoff at 665 nm, and the red cutoff is provided by the Kodak I-N emulsion around 900 nm. The plates taken in the years 1980 and following were taken at the Catania Observatory with the 40/50/120 cm Schmidt telescope and RG695 + I-N filter/emulsion combination.
According to \citet{Bes79} (his Table 7), the Cousins $I_\mathrm{C}$ filter can be approximated with photographic material with an RG 695 filter (instead of RG5) in combination with Kodak IV-N emulsion, which is the improved finer grain version of the I-N emulsion. The blue cutoff difference between these two filters is 30 nm, while the width at half maximum of the $I_\mathrm{C}$ filter is 110 nm, so that the bandwidth increase is about 27\%.  Given that the spectrum of an M-type star is monotonically decreasing blueward within the $I_\mathrm{C}$ band, the magnitude difference between the two filter/emulsion combinations is comparable to, or lower than, the errors of Maffei's magnitudes. Indeed, \citet{Maf13} did not make any correction when merging the datasets of the two instruments to build their light curves.

Anyway, when Maffei established his magnitude scale in 1967 the Cousins system was still not defined; therefore  he established a comparison sequence on the Asiago plates with an approximated Johnson's $I$ magnitude derived by extrapolating the $U\!BV$ magnitudes of 60 stars of the \citet{Wal61} sequence in the open cluster NGC 6611, which is located near the center of Maffei's field. 

We found more details, including the actual list of the stars and their adopted magnitudes, in Montalto's thesis. Instrumental magnitudes for these comparison stars were derived by Maffei with the Becker iris photometer of the Monteporzio (Roma) Observatory and a calibration curve was built using the above mentioned infrared magnitudes. Secondary comparison sequences were then built by Maffei in selected areas of the field of view, to allow the magnitude estimates of the discovered variables, which were made by eye inspection with a microscope. The internal accuracy of Maffei's individual magnitudes is reported to be from 0.1 to 0.3 mag \citep{Maf95}. We stress that no actual infrared standard stars were used by Maffei, so his scale is not linked to the present Cousins system.
 
To perform a transformation of the infrared magnitudes of Maffei's primary sequence ($I_\mathrm{Maf}$) into the OGLE $I_\mathrm{C}$  magnitudes, we recovered from the OGLE archive the $I_\mathrm{C}$ magnitudes of his 60 comparison stars.
The scatter plot of these OGLE magnitudes with Maffei's ones is plotted in Fig.~\ref{fig:mag_calib}, which shows a good linear correlation, namely:
\begin{equation}\label{eq:calib}
I_\mathrm{C}=I_\mathrm{Maf}\times 0.851 (\pm 0.017) + 1.561 (\pm 0.239),
\end{equation}
 with rms deviation 0.21 mag. The slope of the regression line is significantly shallower than 1.0, suggesting the existence of some systematic effect: unfortunately the stars of the Walker sequence do not contain very red stars, so that it is not possible to check the existence of a color term in the transformation from $I_\mathrm{Maf}$ to $I_\mathrm{C}$.

\begin{figure*}
\centering
\includegraphics[width=\linewidth]{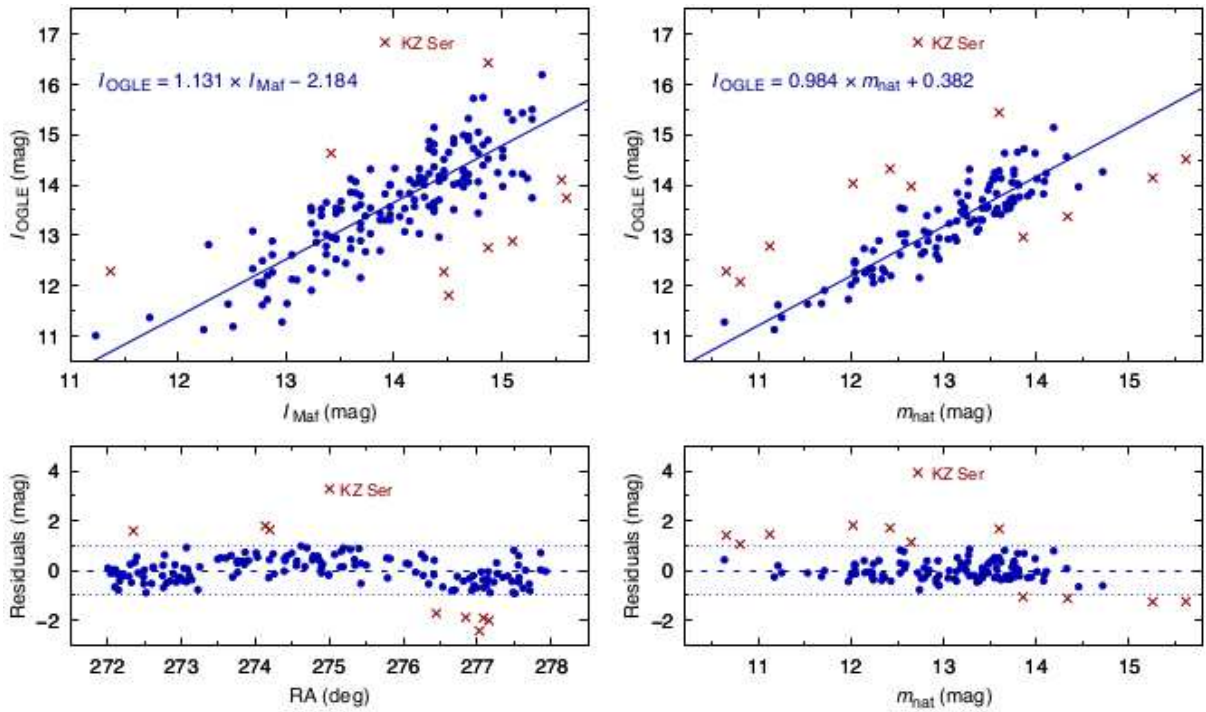} 
\caption{Relations between Maffei's and OGLE magnitudes. \textsl{Top left:} OGLE magnitudes of our variables as a function of the mean Maffei's magnitudes transformed into $I_\mathrm{C}$ according to Eq.~\ref{eq:calib} . \textsl{Bottom left:} residuals of the linear fit as a function of right ascension. \textsl{Top right:} $I_{\mathrm{OGLE}}$ as a function of mean magnitudes derived from the reanalysed light curves. \textsl{Bottom right:} residuals of the linear fit.}
\label{fig:mags}
\end{figure*}

To compare Maffei's magnitudes of the variable stars with the OGLE $I_\mathrm{C}$ ones, we transformed the original Maffei maximum and minimum magnitude of each variable star into $I_\mathrm{C}$ according to Eq.~\ref{eq:calib} and computed the average magnitude ($(\mathrm{max}+\mathrm{min})/2$).
These average Maffei magnitudes are compared with the OGLE mean magnitudes (derived from the fit of the light curve) in the top-left panel of Fig.~\ref{fig:mags}. Excluding the 10 most discrepant stars, the best fitting slope is 1.23, significantly different from the ideal case of 1.00, with an offset of about 0.5 mag at the plot center; remarkably, the spread is quite large, with an rms deviation $\sigma=0.72\,\mathrm{mag}$ from the regression line.

The deviation of the slope from the ideal case, and the offset, might be due to a color term, which we could not evaluate given that the Walker sequence does not contain M-type stars, but it is not a concern to detect a possible large change of the mean magnitude of a given variable at $\sim$50 years distance. More of concern is the large dispersion of the data from the regression line. 

A further source of systematic error may be the way of computing the average value: for the OGLE data it comes from the fitting of a sine curve to the whole light curve, for Maffei's data it comes from a simple average of the maximum and minimum reported values. 

To understand the reason of the large spread we plotted the magnitude differences as a function of the celestial coordinates, finding a significant trend in RA (see Fig.~\ref{fig:mags}, bottom left). A possible reason of this trend could be inconsistencies among the several secondary comparison sequences actually used by Maffei for his magnitude eye-estimates. For instance, the four outlying stars around coordinates (l=277, b=-2, see bottom left panel of Fig.~\ref{fig:mags}) have similar declination, so are in the same sky area, suggesting a systematic error in the local comparison sequence.

\begin{figure}
\centering
\includegraphics[width=10cm]{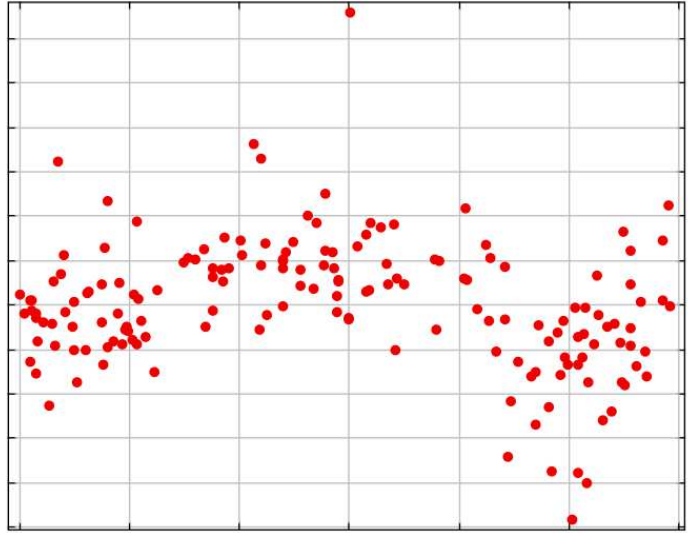} 
\caption{The distance of the OGLE mean magnitude from the regression line of Fig.6 {\it vs} the R.A. of the star. A sistematic trend is present. The very discrepant star at the top is KZ Ser.}
\label{fig7}
\end{figure}

\subsection{Reanalysis of photographic plates}\label{sec:reanalysis}

To minimize the systematic errors expected from the inconsistencies of the secondary comparison sequences and from the way of computing the average magnitudes of the variable stars, we reanalyzed 47 Maffei's plates, obtained with the Asiago 65/90\,cm Schmidt telescope, and digitized at 1600 dpi (1.59 arcsec/pixel). The astrometric solutions and photometric reductions were carried out using the PyPlate software, in a pipeline developed for the APPLAUSE project \citep{Tuv14}. 

The pipeline processed the digitized plate images in multiple steps. Instrumental magnitudes (MAG\_AUTO) of all detected sources were extracted with the SExtractor program \citep{Ber96}. Initial astrometric solutions were derived with the Astrometry.net software \citep{Lang10} and refined astrometric calibration in sub-fields was obtained with SCAMP \citep{Ber06}, using reference coordinates from the UCAC4 catalog. The source list was then cross-matched with the UCAC4 and APASS DR9 catalogs. The typical internal error of the individual PyPlate measures was 0.15 mag, quite comparable to that of the original Maffei's measures, and was basically dictated by the quality of the hypersensitized I-N plates, but in some cases it was up to 0.3 mag. Further details of the procedure can be found in \citet{NTR18}.

Photometric calibration was carried out with the Sloan $r'$ and $i'$ magnitudes of a large number of non variable stars taken from the APASS DR9. Each photographic plate has a unique color response that can be characterized with the color term $C$, as explained by \citet{Lay10}. Magnitudes in the plate natural system, $m_\mathrm{nat}$, are related to the standard magnitudes:
\begin{equation}\label{eq:color_term}
    m_\mathrm{nat} = i' + C(r'-i').
\end{equation}
We determined the color term for each plate by trying a series of $C$ values to minimize the scatter in the photometric calibration curve. The color terms varied from $-0.26$ to $-0.04$, with a mean value of $-0.15$. All instrumental magnitudes were then transformed to natural magnitudes: these $m_\mathrm{nat}$ are our best estimates of the Sloan $i'$ magnitudes of the individual observations of our variables. Conversion into the $I_\mathrm{C}$ band could in principle be made using the formulas by e.g.
\citet{Jor06} or \citet{Jes05}, but these formulas contain color indexes, which are not known for our variables. The expected approximate zero point difference can be evaluated comparing the $m_\mathrm{nat}$ magnitudes of the stars in the Walker sequence with their $I_\mathrm{C}$ magnitudes in the OGLE database: the comparison was quite satisfactory, with
\begin{equation}\label{eq:pyplate_term}
I_\mathrm{C} = m_\mathrm{nat} \times 1.008(\pm0.005)-0.353(\pm0.081).
\end{equation}
At first order therefore an offset of $-0.35$ mag should be expected. As told above in Section \ref{sec:comp-mags}, the Walker sequence does not contain M-type stars, so possible further color term corrections for the magnitudes of our (very red) variables could be present but cannot be quantified.

Unfortunately PyPlate was in trouble for star images near the plate limit, so for many variable stars our light curve is not well populated. Anyway we were able for 128 stars to compute the mean magnitude and the variation amplitude from the sinusoidal fit of their light curve, derived in the same way as for the OGLE light curves described above in Section~\ref{sec:det-per}. Despite the actual light curves of our stars are not strictly sinusoidal, the use of the same shape for the OGLE and the Asiago data minimize systematic differences.
For stars detected only during the high state, the light curve is not constrained in the faint part, so the mean magnitude and amplitude derived from the sinusoidal fit are less reliable.

For the 118 best sampled stars (good number of points well distributed in phase), the mean magnitude is compared with the OGLE one in the top-right panel of Fig.~\ref{fig:mags}.
The regression line shown has slope near unity and the magnitudes offset is negligible.
The spread of the stars around the regression line ($\sigma=0.61$) is reduced with respect to the plot of Fig.~\ref{fig:mags} (left), made using the original Maffei's data. This is an indication that the systematic errors due to the secondary Maffei's comparison sequences, and to the way of computing the average magnitudes, have been reduced, but remains significantly larger than the nominal accuracy of the individual measures.

The most distant stars from the best fit locus are marked in Fig.~\ref{fig:mags} and listed in Table \ref{tab4} in order of mean magnitude difference.

\begin{table*}[t!]
\caption{Stars with possible change of the average magnitude.}\vspace{3mm}
\centering
\begin{tabular}{l l l r r r l}
\hline
Name    & OGLE & Maffei & $I_\mathrm{OGLE}$ & Period & $\Delta I$ & N. of   \\
SIMBAD  &id    & id      & mag              & d      & mag        &  plates \\
\hline
\hline
KZ Ser    &OGLE-BLG-LPV-261988 & M140& 16.84& 511.9& 3.94 & 13 \\
V414 Sct  &OGLE-BLG-LPV-264442 & M137& 15.44& 546.7& 1.67 & 34\\
V3924 Sgr &OGLE-BLG-LPV-261845 & M096& 14.32& 362.8& 1.71 & 14\\
V374 Sct  &BLG787.23.49        & M203& 14.03& 421.4& 1.82 & 23\\
V3927 Sgr &OGLE-BLG-LPV-261965 & M097& 12.78& 218.4  & 1.45 & 38\\
GR Ser    &BLG767.28.8         & M164& 12.28& 234.9& 1.41 & 45\\
V383 Sct  &OGLE-BLG-LPV-263764 & M061& 13.97& 295.5& 1.14 & 33\\
V3947 Sgr &OGLE-BLG-LPV-263995 & M060& 12.07& 308.1 & 1.06 & 41\\
V386 Sct  &OGLE-BLG-LPV-263898 & M026& 12.86& 472.4  & -1.06& 29\\
V398 Sct  &OGLE-BLG-LPV-264150 & M145& 13.37& 183.3  & -1.12& 38\\
GS Ser    &OGLE-BLG-LPV-259041 & M020& 14.51& 373.9 & -1.24& 25\\
\hline
\end{tabular}\label{tab4}
\end{table*} 

We have checked the light curves of these stars to see if an uneven sampling could produce a strong bias. The stars above the line (positive delta mag) are all detected only around the tip of the light curve from the Pyplate pipeline, so their computed average values are unreliable. The stars below the line (brighter than OGLE in the Asiago plates) are poorly sampled with the plates available to us. After careful examination of the light curves we conclude that only two stars have a reliable difference in their mean magnitude, KZ Ser and GR Ser.

The most striking case (4 mag deviation) is that of KZ Ser. This Mira variable has one of the longest period of this sample: the variability amplitude (about 3 mag) is very similar in both datasets, as well as the shape of the light curve, with a very steep rising branch and a long decreasing branch. To check our result we found in the STScI archive two digitized infrared plates (N emulsion) of the Second Palomar Sky Survey (POSS-II) survey, taken on 1980-08-29 (MJD 44480) and 1982-04-18 (MJD 45077):  the first plate is simultaneous to a  Maffei's plate taken at the Catania Observatory, while there are no plates around the second date. 

At the first POSS-II plate epoch the star was near its minimum (phase 0.77) according to Maffei's ephemeris: \citet{Maf13} report a magnitude $I_\mathrm{Maf}$=16.0 which is converted to  $I_\mathrm{C}$=15.2 according to Eq.~\ref{eq:calib}), in fair agreement with the GSC2.3 catalog (15.8). 
On the second POSS-II plate the star was at phase 0.92, on the steep rising branch, so its expected magnitude is rather uncertain. On the POSS-II plate the star was about 15.4, somewhat brighter than in the first plate, in fair agreement with the expectation. The POSS-II plates therefore confirm Maffei's observations, so we think that the luminosity drop of $\sim$4 mag of KZ Ser detected by OGLE is real.
A possible explanation for the star fading may be extinction by a recent dust emission.

GR Ser had a brighter average magnitude at the time of Maffei's monitoring. It is covered by only one public POSS-II plate (taken on 1980-08-29), simultaneous to a Maffei's one, with  $I_\mathrm{C}$=10.88 in the GSC2.3 catalog. Maffei's value is 11.3 (rescaled as above), in fair agreement. 
The OGLE magnitudes range from $I_\mathrm{C}$ 9.5 to 16, while Maffei's monitoring ranged from 10.1 to 13.0 (rescaled according to Eq.~\ref{eq:calib}). Overall the maxima are fairly consistent, so we guess that the different average value is due to the deep minimum detected by OGLE and missed by Maffei.

\section{Infrared colors and distance estimates}\label{sec:dist}

Our variables are located around galactic coordinates $l=16^\circ$, $b=0^\circ$. From the map of the galactic plane by \citet{Reid14} we expect that they belong to the Scutum and the Sagittarius arms so may be largely spread in distance form 1 to 4 kpc. They may therefore suffer of different absorptions, of the order of 1 to 3 mag in $V$, or even more.

\begin{figure}
\centering
\includegraphics[width=\columnwidth]{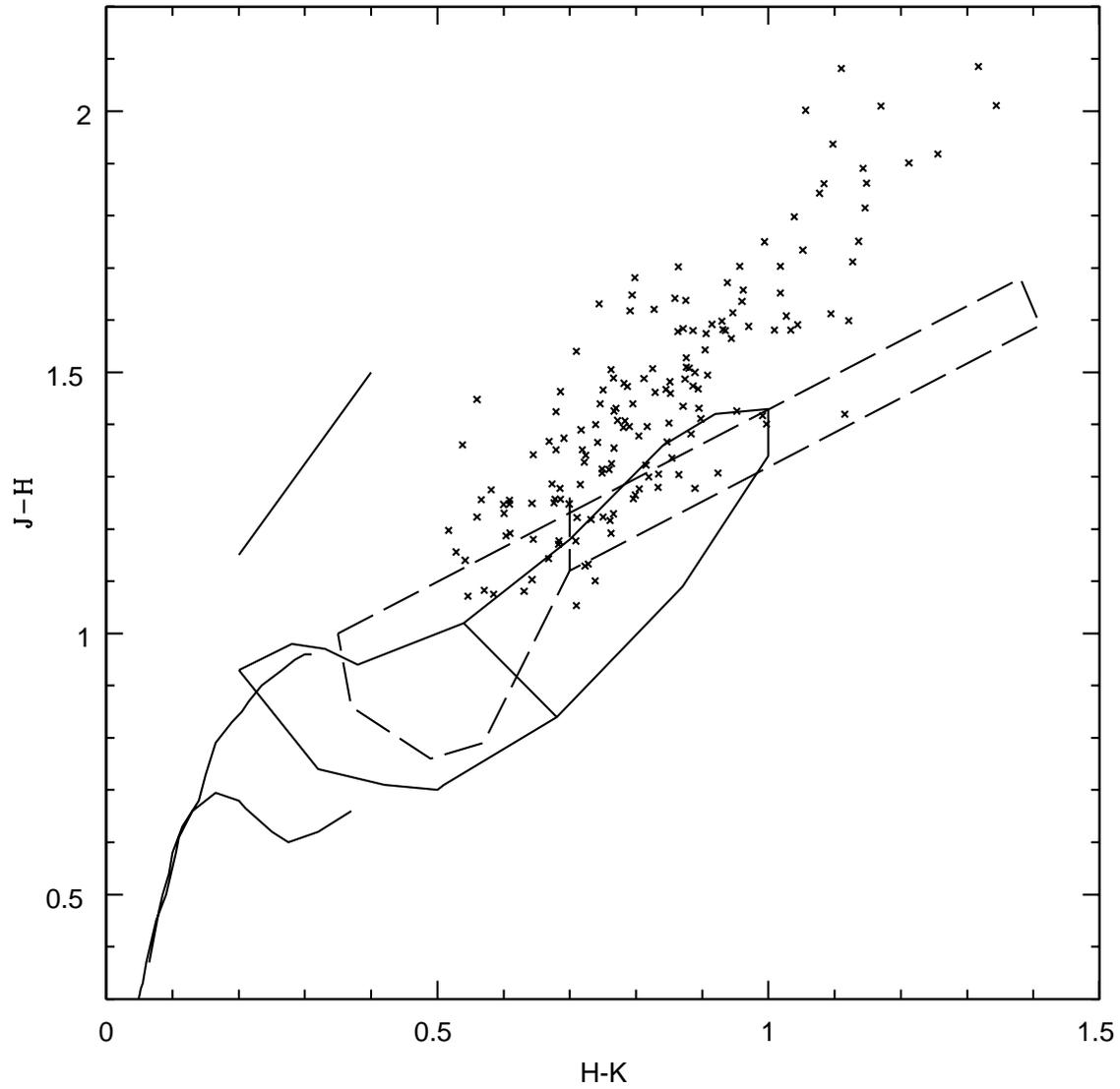} 
\caption{The NIR color-color plot of Maffei's LPVs with an OGLE period overplotted to the Bessell \& Brett loci for Late Type Stars described in the text. The reddening vector for $E(B-V)=1$ is drawn to help the eye.  }
\label{fig:color-color}
\end{figure}

Having identified a 2MASS counterpart for all our Long Period Variables, we build their $(J-H)$ {\bf vs} $(H-K)$ diagram (see Fig.~\ref{fig:color-color}) where we reported also the loci of Late Type Stars as defined by \citet{BB88} (their Fig.~A3). The 2MASS passbands are not exactly those used by \citet{BB88}, but our purpose is just to check an overall consistency and the presence of significant reddenings. In this plot the continuous line at the bottom left is the Main Sequence, which separates around spectral type M0 into a Red Giants (upper) and a Main Sequence (lower) branch. The area enclosed by hatched lines is mostly populated by variable Carbon stars; that enclosed by the continuous line is populated by Oxigen-rich Long Period Variables. To help the reader, we report in this figure also the reddening vector for $E(B-V)=1$, computed according to \citet{Sch98}.

As expected, all our stars are located in or above the area of the Long Period Variables. Color corrections corresponding to $E(B-V)$ ranging from 1 to  3 mag are required to bring all our stars within the borders of the Oxigen-rich stars locus (continuous line). A discussion of the individual reddening is beyond the purpose of this paper: JHK magnitudes averaged over the full light curve should be necessary, instead of the single snapshots given by the 2MASS catalog.

We searched for a distance estimate of our stars in the Gaia DR2 database \citep{Gaia18} as available at CDS: in all cases a matching source with very red $BP-RP$ color was found within 0.35~arcsec from the 2MASS position. 

Unfortunately, reliable parallaxes are not available for any of our stars, because the parameters indicating the goodness of the astrometric solution (astrometric\_gof\_al, astrometric\_excess\_noise, astrometric\_weight\_al) are all very poor \citep{Gaia-doc}. No parallax accuracy improvements for our stars were given in the EDR3 version \citep{Gaia-CDS}.
The formal median parallax of our sample is 0.15 mas with nearly symmetric and large spread ($\sigma=0.59$ mas), many stars having formally negative parallax. The median proper motions of the sample are -2.26 mas/yr in RA and -4.38 mas/yr in Dec, corresponding to -5.0 mas/yr in Galactic longitude and 0.0 mas/yr in Galactic latitude, in fair agreement with the expectations given their positions in the sky.

Assuming the period-absolute $K$ mag relation of \citet{Whit12} and a color excess $E(B-V)=1$, we get a median distance for the sample of 3.7 kpc, in fair agreement with the expectation. An accurate distance determination for each star would require a knowledge of the individual reddening and of the actual mean $K$ magnitude instead of the 2MASS snapshot value, which is taken at an unknown phase. 

\section{Conclusions}\label{sec:conc}

The large majority of the 150 Long Period Variables in Maffei's M16-M17 sample show a substantially stable period at a distance of 40--50 years. The original classification as Mira or SR is not supported by the extensive work by \citet{Sos13}  on the Red Variables in the Galactic bulge, which sets as discriminating parameter a "peak to peak" amplitude $\Delta I_\mathrm{C} \le 0.8$ mag. 
In this respect nearly all the LPVs of the sample should be considered as Miras, save three (V397 Ser, NSV 10832, and NSV 10772); increasing (somewhat arbitrarily) the limit $\Delta I_\mathrm{C}$ to 1.0~mag brings the potential number of SR to just 8 stars, against 24 of the original classification.

A statistical comparison can be done with the finding by \citet{Tem05}, who studied 547 Mira or possible Mira (M:) stars from the AAVSO database: none of those stars is included in our sample. In the AAVSO sample, only 3 stars showed a period variation larger than 10\%: no surprise that we found no star with a significant period variation in our sample of 150 stars.

Regarding the average magnitudes of the LPVs observed by Maffei, they are well correlated with the OGLE mean magnitudes, with a linear relation of slope nearly 1. However, the spread around this relation is quite large ($\sigma=0.61\,\mathrm{mag}$), substantially larger than the typical photographic photometry error. It is therefore possible that a few stars have actually changed their mean luminosity by 1 to 2 magnitudes: the most relevant case, with a $\sim$4 mag dimming, is KZ Ser. 

During this research we found the right identification for 3 misidentifed Maffei's variables (NSV 10326 alias M005, NSV 10522 alias M035, and IX Ser alias M179), which will allow their study in the future.

Finally, we point out the reclassification as Mira of five stars in Maffei's sample (GL Ser, NSV 10271, NSV 10326, NSV 10522, NSV 10677), not recognized as such in the original discovery papers.

\setcounter{secnumdepth}{0}
\OEJVacknowledgements{
Part of this work was supported by institutional research funding IUT40-2 of the Estonian Ministry of Education and Research. TT acknowledges the support by the Centre of Excellence ''Dark side of the Universe`` (TK133) financed by the European Union through the European Regional Development Fund.
This work has made use of data from the European Space Agency (ESA) mission
{\it Gaia} (\url{https://www.cosmos.esa.int/gaia}), processed by the {\it Gaia} Data Processing and Analysis Consortium (DPAC,
\url{https://www.cosmos.esa.int/web/gaia/dpac/consortium}). Funding for the DPAC has been provided by national institutions, in particular the institutions participating in the {\it Gaia} Multilateral Agreement.
This research has made use of the WEBDA database, operated at the Department of Theoretical Physics and Astrophysics of the Masaryk University. We thank dr. Corinne Rossi for helpful suggestions.}
\\

\begin{verbatim}
SIMBAD       2MASS           OCVS         Maffei Period err T0 Magmax Magmin
FZ_Ser 18080193-1444151 OGLE-BLG-LPV-258610 M173 243.1 0.4 56939 12.5 15.7
GG_Ser 18081103-1434279 OGLE-BLG-LPV-258664 M101 362.2 0.8 56879 10.8 12.2
GH_Ser 18082202-1524048 OGLE-BLG-LPV-258711 M030 276.3 0.5 56917 12.1 14.2
GI_Ser 18082639-1535119 OGLE-BLG-LPV-258733 M090 252.1 0.5 56963 11.2 14.1
GK_Ser 18082549-1418077 OGLE-BLG-LPV-258729 M103 373.1 0.4 57123 13.6 16.1
GL_Ser 18083439-1507387 OGLE-BLG-LPV-258777 M031 171.2 0.2 56963 12.6 15.8
GM_Ser 18083581-1504017 BLG766.05.11 M092 193.8 0.8 57006 10.6 11.7
GN_Ser 18084003-1509393 OGLE-BLG-LPV-258806 M032 273.4 0.4 57060 12.0 15.3
GO_Ser 18085193-1408331 OGLE-BLG-LPV-258863 M003 237.3 0.4 57028 12.3 15.2
GP_Ser 18090980-1551201 OGLE-BLG-LPV-258945 M163 288.2 0.4 57058 12.1 14.9
GQ_Ser 18091802-1438126 OGLE-BLG-LPV-258994 M100 490.5 0.6 56969 12.5 15.8
GR_Ser 18092439-1519364 BLG767.28.8 M164 234.9 0.7 56950 9.4 16.0
GS_Ser 18092910-1407529 OGLE-BLG-LPV-259041 M020 373.9 0.3 57138 13.4 15.8
GT_Ser 18093492-1426408 OGLE-BLG-LPV-259071 M104 330.1 1.2 57119 13.3 16.6
GU_Ser 18093953-1455390 OGLE-BLG-LPV-259090 M002 330.2 1.5 57083 12.6 16.1
GV_Ser 18095709-1420456 OGLE-BLG-LPV-259169 M019 237.0 0.2 57002 13.4 15.3
GW_Ser 18095980-1451159 OGLE-BLG-LPV-259183 M105 296.8 0.5 56984 11.9 14.0
GX_Ser 18100565-1514019 OGLE-BLG-LPV-259204 M001 190.0 0.4 57016 10.6 13.1
GZ_Ser 18102342-1522234 OGLE-BLG-LPV-259281 M033 329.6 1.0 56885 10.9 13.9
HH_Ser 18102941-1507575 OGLE-BLG-LPV-259302 M042 490.3 1.0 57332 11.5 14.8
HI_Ser 18102813-1426061 OGLE-BLG-LPV-259297 M177 120.5 0.1 56869 12.3 15.7
HK_Ser 18104075-1352176 BLG766.27.8933 M050 446.0 0.0 40464 12.5 13.9
HL_Ser 18105937-1426022 OGLE-BLG-LPV-259454 M004 369.4 0.7 57159 12.8 14.9
HM_Ser 18110188-1535566 OGLE-BLG-LPV-259470 M034 197.5 0.2 57012 11.7 13.9
HN_Ser 18111261-1545228 OGLE-BLG-LPV-259535 M167 333.2 2.4 56942 12.7 16.3
HP_Ser 18112364-1537321 OGLE-BLG-LPV-259581 M093 478.1 1.1 57079 12.2 15.2
HQ_Ser 18113340-1544212 OGLE-BLG-LPV-259643 M094 314.1 0.7 57125 12.7 15.8
HR_Ser 18113915-1447054 OGLE-BLG-LPV-259675 M106 429.4 2.1 57288 11.2 14.0
HS_Ser 18114344-1508563 OGLE-BLG-LPV-259699 M017 516.8 0.8 57340 10.8 13.9
HT_Ser 18114978-1537187 OGLE-BLG-LPV-259738 M098 472.2 2.0 57287 12.2 15.1
HU_Ser 18115480-1540129 OGLE-BLG-LPV-259767 M095 401.0 1.5 57014 11.8 14.9
HV_Ser 18115776-1539244 OGLE-BLG-LPV-259785 M040 523.4 1.4 57316 11.5 14.2
HW_Ser 18120707-1454509 OGLE-BLG-LPV-259830 M045 286.9 0.6 57104 11.1 14.4
HX_Ser 18120879-1444481 OGLE-BLG-LPV-259839 M049 215.6 0.5 56942 13.3 16.1
HY_Ser 18121704-1547221 OGLE-BLG-LPV-259880 M039 204.0 0.2 57011 11.2 13.9
HZ_Ser 18121649-1424381 OGLE-BLG-LPV-259874 M008 454.5 1.0 56959 12.0 15.7
II_Ser 18122008-1522443 OGLE-BLG-LPV-259900 M041 328.7 0.8 56928 12.8 15.5
IK_Ser 18122562-1545195 OGLE-BLG-LPV-259930 M038 335.4 0.7 56886 12.1 14.8
IL_Ser 18123403-1520574 OGLE-BLG-LPV-259965 M107 202.9 0.3 56894 13.1 15.1
IM_Ser 18125363-1446150 OGLE-BLG-LPV-260070 M176 195.0 0.4 57027 12.7 14.3
IN_Ser 18125971-1454306 OGLE-BLG-LPV-260102 M207 474.4 0.8 57194 12.4 15.9
IO_Ser 18142466-1531035 OGLE-BLG-LPV-260511 M018 386.9 0.9 57026 11.7 14.0
IP_Ser 18144622-1407105 OGLE-BLG-LPV-260602 M117 214.1 0.8 56970 10.7 13.5
IQ_Ser 18150079-1548166 OGLE-BLG-LPV-260662 M006 356.7 0.7 57023 14.5 17.1
IR_Ser 18150239-1433461 OGLE-BLG-LPV-260675 M056 192.4 0.4 56997 12.4 14.9
IS_Ser 18150248-1407502 OGLE-BLG-LPV-260676 M022 436.9 1.2 57074 11.9 14.7
IT_Ser 18151980-1308503 OGLE-BLG-LPV-260767 M062 522.1 1.7 57396 12.5 15.6
IU_Ser 18152445-1402071 OGLE-BLG-LPV-260784 M181 372.2 1.1 57196 12.9 15.9
IV_Ser 18152657-1250317 OGLE-BLG-LPV-260794 M131 374.5 0.7 57202 13.2 15.8
IX_Ser 18153810-1505332 OGLE-BLG-LPV-260841 M179 301.9 0.8 56930 13.8 17.7
IY_Ser 18160250-1518028 OGLE-BLG-LPV-260944 M180 465.0 0.3 57279 12.3 15.0
IZ_Ser 18160690-1458486 OGLE-BLG-LPV-260964 M178 314.8 1.1 57168 13.5 16.3
KK_Ser 18190511-1243207 OGLE-BLG-LPV-261762 M013 287.0 0.7 57004 13.0 16.0
KM_Ser 18164727-1314374 OGLE-BLG-LPV-261136 M129 461.0 0.9 57229 12.7 16.0
KN_Ser 18165790-1553252 OGLE-BLG-LPV-261177 M110 350.8 0.5 57142 13.0 15.7
KO_Ser 18170085-1518195 OGLE-BLG-LPV-261194 M007 286.5 0.8 56934 10.5 13.6
KP_Ser 18173587-1239199 OGLE-BLG-LPV-261359 M194 242.6 0.8 56997 12.3 15.2
KQ_Ser 18175796-1304519 OGLE-BLG-LPV-261483 M189 294.7 0.4 57150 13.2 16.0
KR_Ser 18181305-1210243 OGLE-BLG-LPV-261546 M141 395.9 0.4 57046 12.1 14.0
KS_Ser 18181494-1227070 OGLE-BLG-LPV-261558 M142 411.5 0.6 56960 12.1 14.6
KT_Ser 18184060-1444469 OGLE-BLG-LPV-261660 M122 389.1 0.7 56928 10.8 13.6
KU_Ser 18190828-1447427 OGLE-BLG-LPV-261772 M009 451.1 3.0 56993 12.3 15.9
KV_Ser 18192266-1223526 OGLE-BLG-LPV-261824 M143 418.9 0.7 57237 12.4 14.8
KW_Ser 18193642-1439294 OGLE-BLG-LPV-261879 M121 223.3 0.4 57039 10.8 12.1
KX_Ser 18193773-1234294 OGLE-BLG-LPV-261886 M144 434.8 1.9 56939 12.4 15.8
KY_Ser 18195866-1236145 OGLE-BLG-LPV-261977 M081 324.3 1.5 56885 11.8 14.5
KZ_Ser 18200040-1309422 OGLE-BLG-LPV-261988 M140 511.9 2.5 57302 15.2 18.4
LM_Ser 18201740-1309171 OGLE-BLG-LPV-262058 M139 419.8 0.6 56936 13.6 16.1
LN_Ser 18203823-1332474 OGLE-BLG-LPV-262154 M190 189.7 0.4 57051 12.6 14.5
LO_Ser 18203583-1208358 OGLE-BLG-LPV-262143 M082 305.9 0.7 56925 12.3 15.6
LP_Ser 18204489-1228399 OGLE-BLG-LPV-262177 M148 256.8 1.1 57065 12.6 15.5
LQ_Ser 18204589-1252469 OGLE-BLG-LPV-262180 M197 394.4 0.5 57103 12.7 15.8
LR_Ser 18212341-1541420 OGLE-BLG-LPV-262341 M118 240.4 0.4 57059 12.9 15.8
NSV_10249 18082415-1535166 OGLE-BLG-LPV-258722 M091 191.4 0.2 56956 13.4 15.0
NSV_10271 18091451-1429483 OGLE-BLG-LPV-258970 M174 300.3 1.0 57274 13.1 16.8
NSV_10326 18110050-1423082 BLG766.18.154 M005 314.8 0.9 57130 14.1 16.3
NSV_10394 18140763-1506343 OGLE-BLG-LPV-260424 M021 310.0 0.8 57162 13.8 16.4
NSV_10408 18144139-1503536 OGLE-BLG-LPV-260585 M051 294.0 0.3 57105 13.7 16.4
NSV_10497 18173619-1502250 BLG779.06.3 M114 4.1 0.0 0 10.7 12.1
NSV_10522 18182915-1725379 OGLE-BLG-LPV-261605 M035 455.8 0.7 0 14.2 18.6
NSV_10677 18213621-1242312 OGLE-BLG-LPV-262404 M028 393.8 0.9 56956 14.0 16.6
NSV_10681 18214295-1300424 BLG778.17.9009 M159 1.8 0.0 0 13.4 14.8
NSV_10693 18215979-1634071 BLG569.16.43 M111 0.0 0.0 0 14.1 15.6
NSV_10772 18254743-1611475 BLG569.19.104 M183 262.2 1.3 56910 13.5 13.9
NSV_10832 18274962-1343087 BLG574.06.26871 M087 326.7 1.5 57071 11.3 12.5
NSV_10899 18295552-1518384 OGLE-BLG-T2CEP-1311 M024 68.1 0.0 0 12.6 13.8
V374_Sct 18230920-1237573 BLG787.23.49 M203 421.4 1.6 57028 13.4 14.8
V375_Sct 18231684-1301356 BLG787.06.9360 M147 512.3 4.0 57330 13.3 16.3
V377_Sct 18245921-1418133 OGLE-BLG-LPV-263346 M071 514.3 1.6 56941 11.7 16.3
V379_Sct 18252234-1531597 OGLE-BLG-LPV-263451 M010 412.4 0.9 57028 11.3 14.6
V380_Sct 18254033-1443006 OGLE-BLG-LPV-263528 M138 443.9 1.8 57086 12.1 15.0
V381_Sct 18255250-1532074 BLG789.01.34839 M188 395.8 1.4 57054 12.6 14.5
V382_Sct 18260817-1447151 OGLE-BLG-LPV-263639 M069 563.6 1.2 57463 10.6 13.3
V383_Sct 18263656-1550387 OGLE-BLG-LPV-263764 M061 295.5 0.3 57068 12.6 15.3
V384_Sct 18265495-1540225 OGLE-BLG-LPV-263846 M128 461.7 0.6 57027 12.8 16.2
V386_Sct 18271563-1549354 OGLE-BLG-LPV-263898 M026 472.4 1.0 57204 11.5 14.5
V387_Sct 18271547-1532243 OGLE-BLG-LPV-263897 M063 387.0 1.8 57179 12.2 14.8
V388_Sct 18273471-1338110 OGLE-BLG-LPV-263951 M204 372.6 0.5 44374 10.8 14.2
V389_Ser 18083617-1447341 OGLE-BLG-LPV-258786 M172 236.9 0.4 57097 13.2 16.0
V390_Ser 18090604-1518371 OGLE-BLG-LPV-258926 M165 170.8 0.6 57018 11.5 13.2
V3904_Sgr 18110608-1613039 OGLE-BLG-LPV-259494 M089 356.8 0.7 57015 11.4 15.6
V3905_Sgr 18111338-1603290 OGLE-BLG-LPV-259541 M162 319.9 0.4 56911 13.3 19.7
V391_Sct 18280661-1554440 BLG576.14.26 M064 0.0 0.0 0 11.7 15.7
V391_Ser 18095802-1458356 OGLE-BLG-LPV-259174 M099 245.1 0.3 57022 12.0 15.1
V3918_Sgr 18290441-1353350 OGLE-BLG-LPV-264248 M086 401.8 1.2 57027 11.3 14.4
V392_Sct 18281253-1514385 OGLE-BLG-LPV-264078 M193 484.8 0.7 57346 12.8 15.3
V3920_Sgr 18173387-1732054 OGLE-BLG-LPV-261345 M088 465.5 0.2 57029 12.3 15.6
V3921_Sgr 18173623-1741404 OGLE-BLG-LPV-261362 M029 528.5 1.7 57043 12.8 15.7
V3924_Sgr 18192768-1745477 OGLE-BLG-LPV-261845 M096 362.8 1.2 57238 12.7 14.9
V3925_Sgr 18193309-1746442 OGLE-BLG-LPV-261863 M037 152.1 0.2 56898 11.0 16.2
V3926_Sgr 18193414-1748395 OGLE-BLG-LPV-261868 M168 285.6 0.5 57072 11.7 14.3
V3927_Sgr 18195744-1743099 OGLE-BLG-LPV-261965 M097 218.4 0.3 56895 11.5 14.1
V3930_Sgr 18210828-1706231 OGLE-BLG-LPV-262275 M108 432.3 1.3 57286 13.0 17.1
V3931_Sgr 18212174-1642586 OGLE-BLG-LPV-262339 M048 439.7 0.5 57027 11.7 14.9
V3932_Sgr 18214251-1732069 OGLE-BLG-LPV-262446 M047 204.4 0.2 56971 11.2 13.4
V3934_Sgr 18230629-1721099 OGLE-BLG-LPV-262824 M109 187.1 0.4 57043 12.8 14.6
V3935_Sgr 18240935-1703399 OGLE-BLG-LPV-263102 M052 403.6 1.0 57185 12.8 15.6
V3936_Sgr 18241836-1707212 OGLE-BLG-LPV-263146 M112 376.1 1.8 57221 12.9 14.3
V3938_Sgr 18243941-1705018 OGLE-BLG-LPV-263241 M054 294.5 0.5 57129 11.5 14.4
V394_Sct 18281864-1525508 OGLE-BLG-LPV-264107 M066 311.2 1.3 57099 11.2 14.7
V394_Ser 18135798-1521596 OGLE-BLG-LPV-260384 M046 326.9 0.4 57174 13.4 16.4
V3940_Sgr 18250931-1636201 OGLE-BLG-LPV-263393 M057 451.5 2.3 57183 10.3 17.8
V3942_Sgr 18253922-1633460 OGLE-BLG-LPV-263526 M120 357.5 2.0 56930 10.1 13.3
V3943_Sgr 18264699-1557304 OGLE-BLG-LPV-263811 M123 367.2 0.7 57005 12.3 15.5
V3944_Sgr 18264776-1621427 OGLE-BLG-LPV-263813 M058 294.4 1.1 56900 12.7 16.0
V3945_Sgr 18272269-1615389 OGLE-BLG-LPV-263921 M023 317.4 1.2 56984 11.1 14.4
V3946_Sgr 18274130-1621546 OGLE-BLG-LPV-263978 M059 405.2 1.4 57264 12.4 14.8
V3947_Sgr 18274739-1607056 OGLE-BLG-LPV-263995 M060 308.1 0.7 56907 10.7 13.7
V3949_Sgr 18281973-1621109 OGLE-BLG-LPV-264113 M126 501.0 2.1 56867 11.2 13.3
V395_Sct 18282052-1506009 OGLE-BLG-LPV-264116 M074 314.1 1.5 56913 11.7 14.5
V3950_Sgr 18283838-1603253 BLG576.05.34937 M127 199.0 0.4 56970 12.3 13.4
V396_Ser 18163239-1242177 OGLE-BLG-LPV-261077 M132 423.9 1.4 56907 14.4 18.4
V397_Sct 18283100-1519244 OGLE-BLG-LPV-264143 M068 318.3 0.6 56881 11.8 15.1
V397_Ser 18164368-1551070 BLG780.16.28126 M175 237.9 0.7 56890 12.5 13.2
V398_Sct 18283358-1456333 OGLE-BLG-LPV-264150 M145 183.3 0.4 56945 12.5 14.4
V398_Ser 18164590-1341174 OGLE-BLG-LPV-261132 M186 319.9 1.5 57022 13.8 16.9
V399_Sct 18283523-1456186 OGLE-BLG-LPV-264162 M075 322.3 1.3 57180 11.9 14.8
V400_Sct 18284236-1522525 OGLE-BLG-LPV-264184 M067 294.1 0.5 57138 11.1 12.5
V400_Ser 18174215-1434567 OGLE-BLG-LPV-261390 M182 426.4 0.7 57133 13.2 16.1
V401_Sct 18285417-1429201 OGLE-BLG-LPV-264225 M080 485.0 1.3 57071 10.7 13.4
V402_Ser 18190905-1242465 OGLE-BLG-LPV-261780 M070 318.4 1.6 56991 14.1 17.4
V403_Sct 18290273-1446582 OGLE-BLG-LPV-264241 M198 333.5 0.8 56956 12.1 14.8
V404_Sct 18291283-1537377 OGLE-BLG-LPV-264281 M065 282.3 0.3 57090 12.6 15.6
V404_Ser 18184746-1237123 OGLE-BLG-LPV-261696 M012 341.6 1.1 57108 12.2 14.4
V405_Sct 18292248-1507598 OGLE-BLG-LPV-264309 M015 355.2 0.7 56879 12.5 15.4
V407_Sct 18293219-1548396 OGLE-BLG-LPV-264335 M135 234.4 0.3 56910 9.9 13.1
V408_Sct 18293898-1446124 BLG575.12.937 M201 321.8 0.0 57357 12.4 15.6
V410_Sct 18295301-1457534 OGLE-BLG-LPV-264383 M199 322.9 1.0 56916 12.4 14.6
V411_Sct 18275904-1343220 OGLE-BLG-LPV-264039 M208 454.0 1.6 57312 11.2 13.4
V412_Sct 18295881-1410102 OGLE-BLG-LPV-264401 M152 412.0 1.5 57019 12.9 15.3
V413_Sct 18300233-1528300 OGLE-BLG-LPV-264415 M136 334.9 1.4 57069 12.7 15.6
V414_Sct 18301384-1527224 OGLE-BLG-LPV-264442 M137 546.7 1.8 41240 13.6 16.8
V415_Sct 18301342-1425191 OGLE-BLG-LPV-264440 M084 302.0 0.8 57092 10.5 14.6
V416_Sct 18301476-1421339 OGLE-BLG-LPV-264443 M205 300.3 0.7 57052 11.1 14.7
V417_Sct 18301572-1431265 OGLE-BLG-LPV-264445 M202 405.5 1.2 57231 10.1 12.8
V418_Sct 18302897-1421353 OGLE-BLG-LPV-264484 M154 399.1 2.9 56871 11.4 13.4
V419_Sct 18303629-1416248 OGLE-BLG-LPV-264499 M153 364.5 0.6 57192 12.1 14.8
V420_Sct 18304816-1452459 OGLE-BLG-LPV-264532 M085 257.0 0.4 57089 10.0 12.5
V421_Sct 18305093-1537289 OGLE-BLG-LPV-264539 M076 218.5 0.8 56959 11.2 14.6
V422_Sct 18312526-1518226 OGLE-BLG-LPV-264637 M078 406.6 1.0 57141 12.3 15.0
V423_Sct 18312523-1443501 OGLE-BLG-LPV-264636 M206 453.4 0.8 56952 12.1 14.7
V424_Sct 18313627-1526581 OGLE-BLG-LPV-264672 M077 473.5 1.0 57244 11.4 14.4
V425_Sct 18314233-1512142 OGLE-BLG-LPV-264693 M200 451.4 3.8 57170 11.3 14.3
V478_Sct 18241283-1315552 OGLE-BLG-LPV-263121 M083 506.5 2.2 57379 13.2 17.4
V5536_Sgr 18250499-1703578 OGLE-BLG-LPV-263376 M113 206.7 0.5 57057 13.1 16.5
\end{verbatim}


\begin{thebibliography}{}
\bibliographystyle{plainnat}

\bibitem[Bertin \& Arnouts(1996)]{Ber96} Bertin, E., \& Arnouts, S. 1996, {\it A\&AS}, {\bf 117}, 393, \OEJVbibcode{1996A\&AS..117..393B}

\bibitem[Bertin(2006)]{Ber06} Bertin, E. 2006, Astronomical Data Analysis Software and Systems XV, 112, \OEJVbibcode{2006ASPC..351..112B}

\bibitem[Bessell(1979)]{Bes79} Bessell, M.~S. 1979, {\it PASP}, {\bf 91}, 589, \OEJVbibcode{1979PASP...91..589B}

\bibitem[Bessell \& Brett(1988)]{BB88} Bessell, M.~S., \& Brett, J.~M. 1988, {\it PASP}, {\bf 100}, 1134, \OEJVbibcode{1988PASP..100.1134B}

\bibitem[Gaia DR2 Documentation(2019)]{Gaia-doc} Gaia DR2 Documentation 2019, 
https://gea.esac.esa.int/archive/documentation/GDR2/

\bibitem[Gaia Collaboration(2018)]{Gaia18} Gaia Collaboration, Brown, A.~G.~A., Vallenari, A., et al. 2018, {\it A\&A} , {\bf 616}, A1, \OEJVbibcode{2018A\&A...616A...1G}

\bibitem[Gaia(2020)]{Gaia-CDS} Gaia EDR3, CDS catalog I/350.

\bibitem[Iwanek et al.(2022)]{Iwa22} Iwanek, P., Soszy{\'n}ski, I., Koz{\l}owski, S., et al. 2022, arXiv:2203.16552, \OEJVbibcode{2022arXiv220316552I}

\bibitem[Jester et al.(2005)]{Jes05} Jester, S., Schneider, D.~P., Richards, G.~T., et al. 2005, {\it AJ}, {\bf 130}, 873, \OEJVbibcode{2005AJ....130..873J}

\bibitem[Jordi et al.(2006)]{Jor06} Jordi, K., Grebel, E.~K., \& Ammon, K. 2006, {\it A\&A}, {\bf 460}, 339, \OEJVbibcode{2006A\&A...460..339J}

\bibitem[Kato(2001)]{Kato01} Kato, T. 2001, Information Bulletin on Variable Stars, 5137, 1, \OEJVbibcode{2001IBVS.5137....1K}

\bibitem[Lang et al.(2010)]{Lang10} Lang, D., Hogg, D.~W., Mierle, K., et al. 2010, {\it AJ}, {\bf 139}, 1782, \OEJVbibcode{2010AJ....139.1782L}

\bibitem[Laycock et al.(2010)]{Lay10} Laycock, S., Tang, S., Grindlay, J., et al. 2010, {\it AJ}, {\bf 140}, 1062, \OEJVbibcode{2010AJ....140.1062L}

\bibitem[Lebzelter(2011)]{Leb11} Lebzelter, T. 2011, {\it Astronomische Nachrichten}, {\bf 332}, 140, \OEJVbibcode{2011AN....332..140L}

\bibitem[Lenz \& Breger(2005)]{Lenz05} Lenz, P., \& Breger, M. 2005, Communications in Asteroseismology, 146, 53, \OEJVbibcode{2005CoAst.146...53L}

\bibitem[Maffei(1975)]{Maf75} Maffei, P. 1975, Information Bulletin on Variable Stars, 985, 1, \OEJVbibcode{1975IBVS..985....1M}

\bibitem[Maffei \& Tosti(1995)]{Maf95} Maffei, P., \& Tosti, G. 1995, {\it AJ}, 109, 2652, \OEJVbibcode{1995AJ....109.2652M}

\bibitem[Maffei \& Tosti(2013)]{Maf13} Maffei, P., \& Tosti, G. 2013, VizieR Online Data Catalog, II/320, \OEJVbibcode{2013yCat.2320....0M}

\bibitem[Moln{\'a}r et al.(2019)]{Molnar19} Moln{\'a}r, L., Joyce, M., \& Kiss, L.~L. 2019, {\it ApJ}, {\bf 879}, 62, \OEJVbibcode{2019ApJ...879...62M}

\bibitem[Nesci et al.(2018)]{NTR18} Nesci, R., Tuvikene, T., Rossi, C., et al. 2018, Revista Mexicana de Astronom\'\i{}a y Astrof\'\i{}sica, 54, 341, \OEJVbibcode{2018RMxAA..54..341N}

\bibitem[Nesci(2018)]{Nes18} Nesci, R.\ 2018, Information Bulletin on Variable Stars, 6255, 1, \OEJVbibcode{2018IBVS.6255....1N}

\bibitem[Paunzen \& Vanmunster(2016)]{Pau16} Paunzen, E. \& Vanmunster, T. 2016, Astronomische Nachrichten, 337, 239, \OEJVbibcode{2016AN....337..239P} 

\bibitem[Press \& Rybicki(1989)]{Press89} Press, W.~H., \& Rybicki, G.~B. 1989, {\it ApJ}, 338, 277, \OEJVbibcode{1989ApJ...338..277P}

\bibitem[Reid et al.(2014)]{Reid14} Reid, M.~J., Menten, K.~M., Brunthaler, A., et al. 2014, {\it ApJ}, {\bf 783}, 130, \OEJVbibcode{1989ApJ...338..277P}

\bibitem[Sabin \& Zijlstra(2006)]{Sabin06} Sabin, L., \& Zijlstra, A.~A. 2006, {\it MemSAIt}, {\bf 77}, 933, \OEJVbibcode{2006MmSAI..77..933S}

\bibitem[Scargle(1982)]{Sca82} Scargle, J.~D. 1982, {\it ApJ}, {\bf 263}, 835, \OEJVbibcode{1982ApJ...263..835S}

\bibitem[Schlegel et al.(1998)]{Sch98} Schlegel, D.~J., Finkbeiner, D.~P., \& Davis, M. 1998, {\it ApJ}, {\bf 500}, 525, \OEJVbibcode{1998ApJ...500..525S}

\bibitem[Soszy{\'n}ski et al.(2011)]{Sos11} Soszy{\'n}ski, I., Udalski, A., Szyma{\'n}ski, M.~K., et al. 2011, {\it AcA}, {\bf 61}, 217, \OEJVbibcode{2011AcA....61..217S}

\bibitem[Soszy{\'n}ski et al.(2009)]{Sos09} Soszy{\'n}ski, I., Udalski, A., Szyma{\'n}ski, M.~K., et al. 2009, {\it AcA}, {\bf 59}, 239, \OEJVbibcode{2009AcA....59..239S}

\bibitem[Soszy{\'n}ski et al.(2013)]{Sos13} Soszy{\'n}ski, I., Udalski, A., Szyma{\'n}ski, M.~K., et al. 2013, {\it AcA}, {\bf 63}, 21, \OEJVbibcode{2013AcA....63...21S}

\bibitem[Schwarzenberg-Czerny(1996)]{SC96}Schwarzenberg-Czerny, A., 1996, {\it ApJ}, {\bf 460}, L107-110, \OEJVbibcode{1996ApJ...460L.107S}

\bibitem[Templeton et al.(2005)]{Tem05} Templeton, M.~R., Mattei, J.~A., \& Willson, L.~A. 2005, {\it AJ}, {\bf 130}, 776, \OEJVbibcode{2005AJ....130..776T}

\bibitem[Tisserand(2012)]{Tis12} Tisserand, P. 2012, {\it A\&A}, {\bf 539}, A51, \OEJVbibcode{2012A\&A...539A..51T}

\bibitem[Tisserand et al.(2013)]{Tis13} Tisserand, P., Clayton, G.~C., Welch, D.~L., et al. 2013, {\it A\&A}, 551, A77, \OEJVbibcode{2013A\&A...551A..77T}

\bibitem[Tuvikene et al.(2014)]{Tuv14} Tuvikene, T., Edelmann, H., Groote, D., et al. 2014, Astroplate 2014, p. 127, \OEJVbibcode{2014aspl.conf..127T}

\bibitem[Udalski et al.(2015)] {Uda15}Udalski, A., Szyma{\'n}ski, M.~K., \& Szyma{\'n}ski, G. 2015, {\it AcA}, 65, 1, \OEJVbibcode{2015AcA....65....1U}

\bibitem[Walker(1961)]{Wal61} Walker, M.~F. 1961, {\it ApJ}, {\bf 133}, 438, \OEJVbibcode{1961ApJ...133..438W}

\bibitem[Whitelock(2012)]{Whit12} Whitelock, P.~A. 2012, {\it Ap\&SS}, {\bf 341}, 123, \OEJVbibcode{2012Ap\&SS.341..123W}

\end{thebibliography}
\end{document}